\documentclass[11pt,a4paper]{scrartcl}

\usepackage{CLICdp}
\usepackage{dirtytalk}
\usepackage{textpos}
\usepackage{CLICdp_definitions}
\usepackage{lineno}
\usepackage{textcomp}

\title{ MEASUREMENT OF THE HIGGS BRANCHING RATIO BR($H\rightarrow\gamma\gamma$) AT 3 TeV CLIC}

\clicdpconf{2021}{002}  

\date{\today}

\addauthor{G. Ka\u{c}arevi\'{c}\thanks{kacarevicgoran@vin.bg.ac.rs}}{\institute{1}}
\addauthor{I. Bo\v{z}ovi\'{c}-Jelisav\u{c}i\'{c}}{\institute{1}}
\addauthor{N. Vuka\u{s}inovi\'{c}}{\institute{1}}

\addauthor{G. Milutinovi\'{c}-Dumbelovi\'{c}}{\institute{1}}

\addauthor{M. Radulovi\'{c}}{\institute{2}}
\addauthor{J. Stevanovi\'{c}}{\institute{2}}
\addauthor{T. Agatonovi\'{c}-Jovin}{\institute{1}}

\addinstitute{1}{Vinca Institute of Nuclear Sciences, University of Belgrade, Belgrade, Serbia}\break
\addinstitute{2}{ University of Kragujevac, Faculty of Science, Kragujevac, Serbia}\break

\onbehalfof{the CLICdp Collaboration} 

\abstract{In this paper we address the potential of a 3 TeV center-of-mass energy Compact Linear Collider (CLIC) to measure the Standard Model (SM) Higgs boson decay to two photons. Since photons are massless, they are not coupled to the Higgs boson at tree level, but they are created in a loop exchange of heavy particles either from the Standard Model or beyond. Any deviation of the effective $H\rightarrow\gamma\gamma$ branching ratio and consequently of the Higgs to photon coupling may indicate New Physics. The Higgs decay to two photons is thus an interesting probe of the Higgs sector, both at the running and future experiments. A similar study has been performed  at 1.4 TeV CLIC, where the statistical uncertainty is determined to be 15\% for an integrated luminosity of 1.5 ab$^{-1}$ with unpolarized beams. 

\noindent This study is performed using a full simulation of the detector for CLIC and by considering all relevant physics and beam-induced processes in a full reconstruction chain. The measurement is simulated on 5000 pseudo-experiments and the relative statistical uncertainty is extracted from the pull distribution.  It is shown that the Higgs production cross-section in $W^+W^-$ fusion times the branching ratio BR($H\rightarrow\gamma\gamma$) can be measured with a relative statistical accuracy of 8.2\%, assuming an integrated luminosity of 5 ab$^{-1}$ with unpolarized beams.  }

\titlecomment{Talk presented at the International Workshop on Future Linear Colliders (LCWS2021), 15-18 March 2021. C21-03-15.1}

\graphicspath{ {./logos/}{./figures/} }

\begin{document}

\titlepage

%
\newcommand{\latex}{\LaTeX\xspace}
\lstset{defaultdialect=[LaTeX]TeX}

\section{Introduction}
\label{sec:Intro}
The Compact Linear Collider (CLIC) is a mature option for a future linear electron-positron collider with TeV scale center-of mass energies. As a Higgs Factory, CLIC offers high-precision measurements of the Higgs boson properties. CLIC is planned to be run in three energy stages, at center-of-mass energies of 380 GeV, 1.4 TeV and 3 TeV, with integrated luminosity of 2  ab$^{-1}$,  2.5 ab$^{-1}$ and 5 ab$^{1}$, respectively. High Higgs production cross-section at higher energy stages enables access to rare decays including $H\rightarrow\gamma\gamma$. The ultimate statistical precision of the Higgs coupling measurements comes from a global fit to many individual measurements across the energy stages. Like other future electron-positron projects, CLIC nicely complements the High-Luminosity-LHC(HL-LHC) physics programme, improving the precision of the Higgs couplings to photons to 1\%  ~\cite{1} (\cref{fig:hllhc}), while the CLIC-only precision of the Higgs coupling to photons is 3.2\% ~\cite{2}. Since photons are massless, coupling of photons to Higgs boson is realized via a loop exchange of heavy particles either from the Standard Model or beyond. Any deviations from the expectation may imply physics beyond Standard Model (SM).  

\section{$BR(H\rightarrow\gamma\gamma)$ measurement at 3 TeV CLIC}
\label{sec:basics}
\begin{figure} [h]
  \begin{minipage}[b]{0.45\textwidth}
    \includegraphics[width=\textwidth]{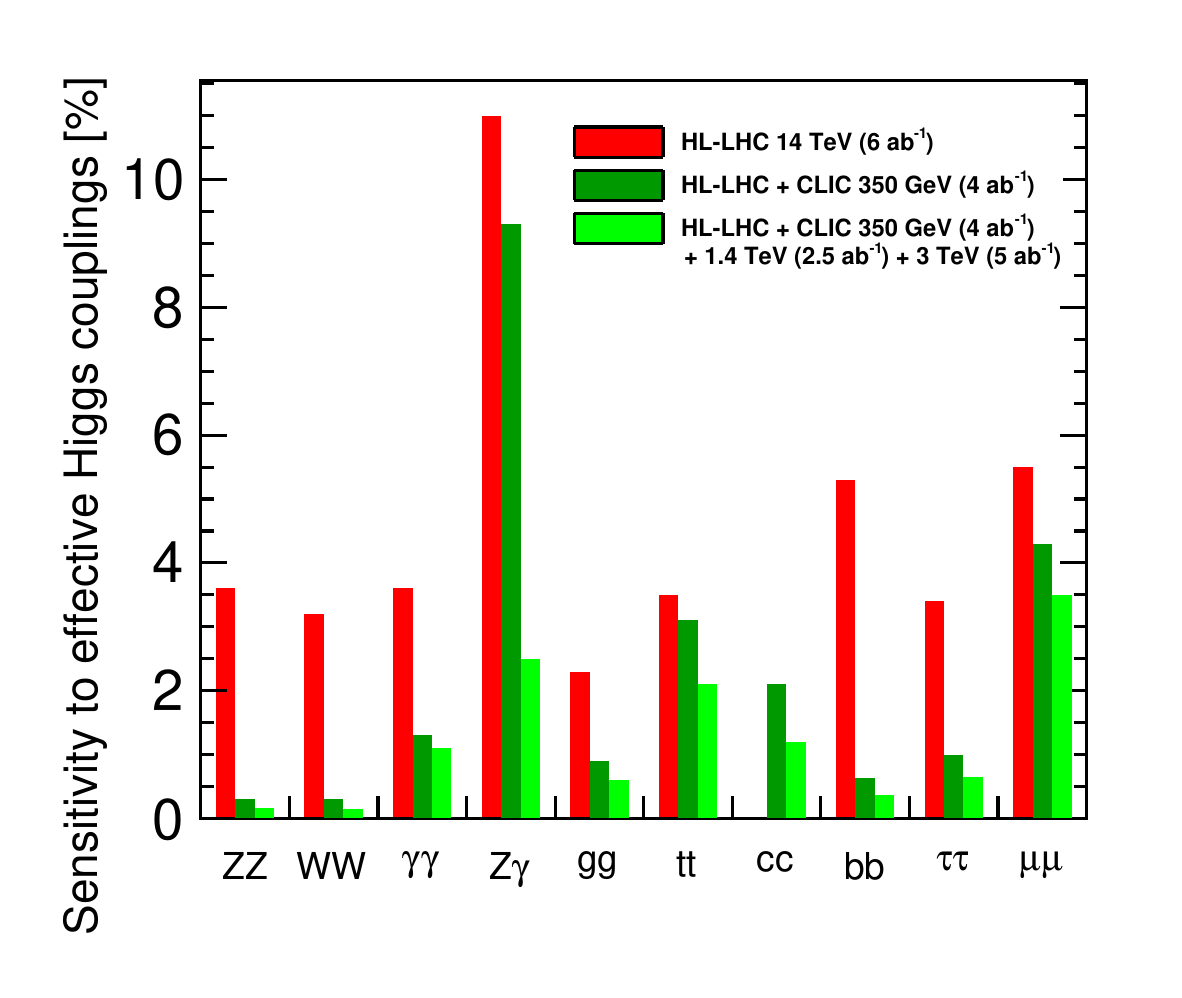}
    \caption{Projections for HL-LHC and HL-LHC + CLIC sensitivity to effective Higgs couplings.}
    \label{fig:hllhc}
  \end{minipage}%
  \hfill 
  \begin{minipage}[b]{0.47\textwidth}
    \includegraphics[width=\textwidth]{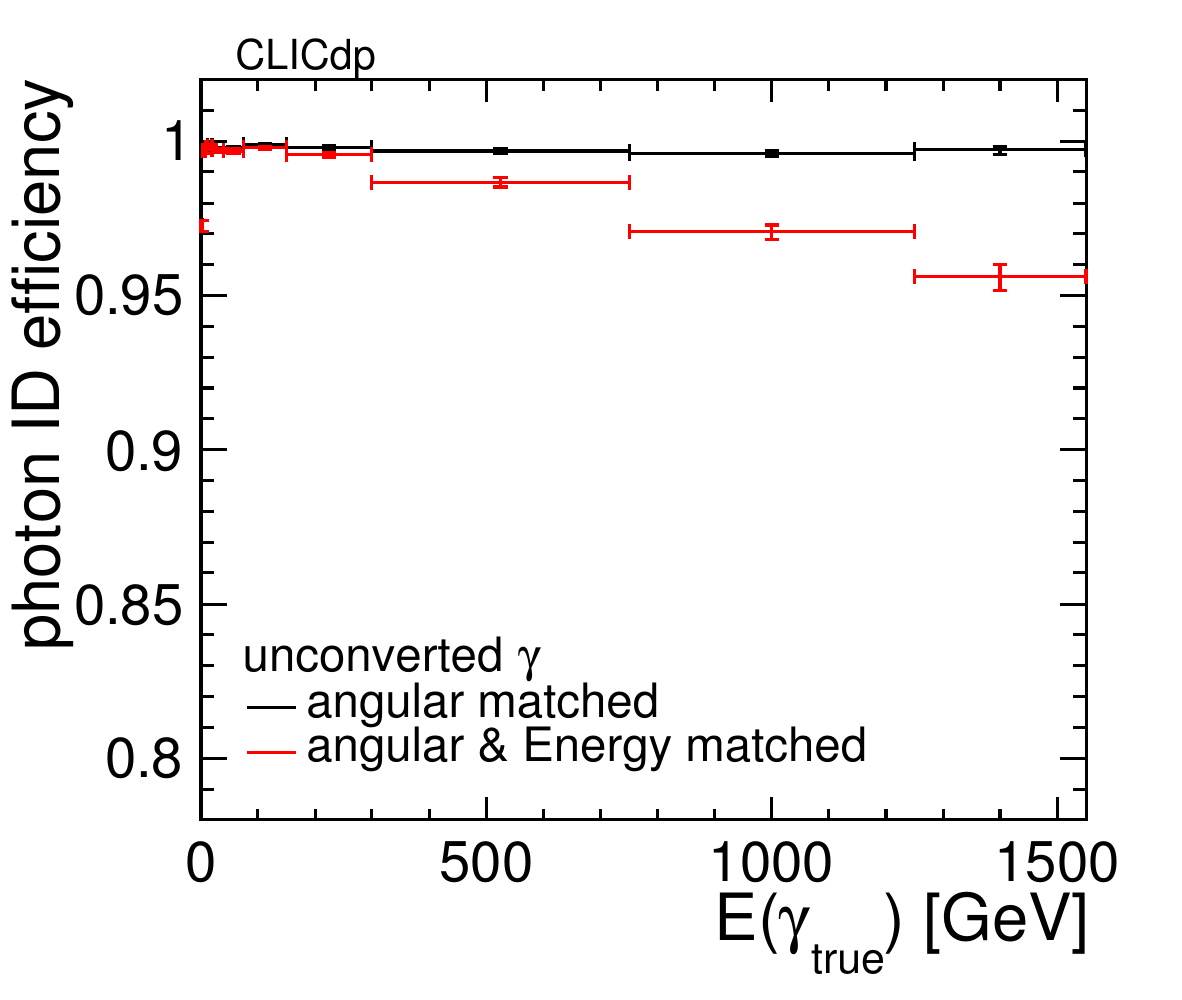}
    \caption{Photon identification efficiency as a function of photon energy.}
    \label{fig:photon_e_res}
  \end{minipage}
\end{figure}

The Higgs production in WW-fusion, as well as background processes are generated in Whizard 1.95 ~\cite{3} including a realistic CLIC luminosity spectrum and beam-induced effects. Interactions with the detector are simulated using the CLIC\_ILD detector model within the Mokka simulation package~\cite{4}. The CLIC\_ILD detector concept is based on fine-grained electromagnetic and hadronic calorimeters optimized for particle-flow techniques (PFA) ~\cite{5}. High granularity in combination with the information from the central tracker within the PFA framework leads to photon reconstruction efficiency of 99\% and electron efficiency of 96\% ~\cite{6}. Photon identification efficiency is illustrated in \cref{fig:photon_e_res} ~\cite{6}.
In electron-positron collisions at 3 TeV center-of-mass energy, the Higgs boson is dominantly produced via WW-fusion with an effective cross-section of 415 fb including ISR effects as well as realistic CLIC luminosity spectrum. The production cross-section can be further increased by electron beam polarization due to a chiral nature of the charged current interaction.

\break
\break
\break

 The branching ratio for Higgs decay to two photons $BR (H\rightarrow\gamma\gamma)$ is 0.23\% which leads to 4750 signal events in 5 ab$^{-1}$ of data. In Table 1, the full list of the signal and considered background processes is given with the corresponding effective\footnote[1]{To reduce simulation time phase, space for di-photon production is restricted to the central tracker region and di-photon mass window around 126 GeV} cross-sections. 
 
 \begin{table}
 \centering
 
 \label{parset}
 \begin{tabular*}{\columnwidth}{@{\extracolsep{\fill}}llll@{}}
 \hline
 \multicolumn{1}{@{}l}{Signal process}  & $\sigma(fb)$  & $N@5 ab^{-1}$  & $N_{simulated}$ \\
 \hline
 $e^+e^-\rightarrow H\nu\nu, H\rightarrow\gamma\gamma$          & 0.95    & 4750  & 24550   \\
 \hline
 \multicolumn{1}{@{}l}{Background processes}  & $\sigma(fb)$ \\
 \hline
 $e^+e^-\rightarrow\gamma\gamma$            & 19      & $9.5\cdot10^5$  &$3\cdot10^4$   \\ 
 $e^+e^-\rightarrow e^+e^-\gamma$           &  797     & $4.0\cdot10^6$ 	& $3\cdot10^6$    \\
 $e^+e^-\rightarrow e^+e^-\gamma\gamma$            &  56   & $2.8\cdot10^5$ 	 &$1.5\cdot10^5$          \\
 $e^+e^-\rightarrow\nu\bar{\nu}\gamma$           &  47   & $2.4\cdot10^4$	&$2\cdot10^5$             \\
 $e^+e^-\rightarrow\nu\bar{\nu}\gamma\gamma$            &  49 & $2.5\cdot10^5$		& $1.6\cdot10^5$         \\
 $e^+e^-\rightarrow q\bar{q}\gamma$             & 584     & $3.0\cdot10^6$   & $1.2\cdot10^6$         \\
 $e^+e^-\rightarrow q\bar{q}\gamma\gamma$            & 72      &$3.7\cdot10^5$  & $3\cdot10^5$      \\
 
 \hline
 
 \end{tabular*}
 \caption{\label{tab-lop}Signal and background considered processes with the corresponding cross-sections at 3 TeV centre-of-mass energy. All processes are produced with generator level cuts to reduce CPU time, requiring, among the others, 100 GeV < $M_{\gamma\gamma}$ < 150 GeV mass window for two-photon system. } 
 \end{table}
 
\subsection{Event selection}
\label{sec:basics}
\begin{figure}[h!]
\centering
\includegraphics[width=0.6\textwidth]{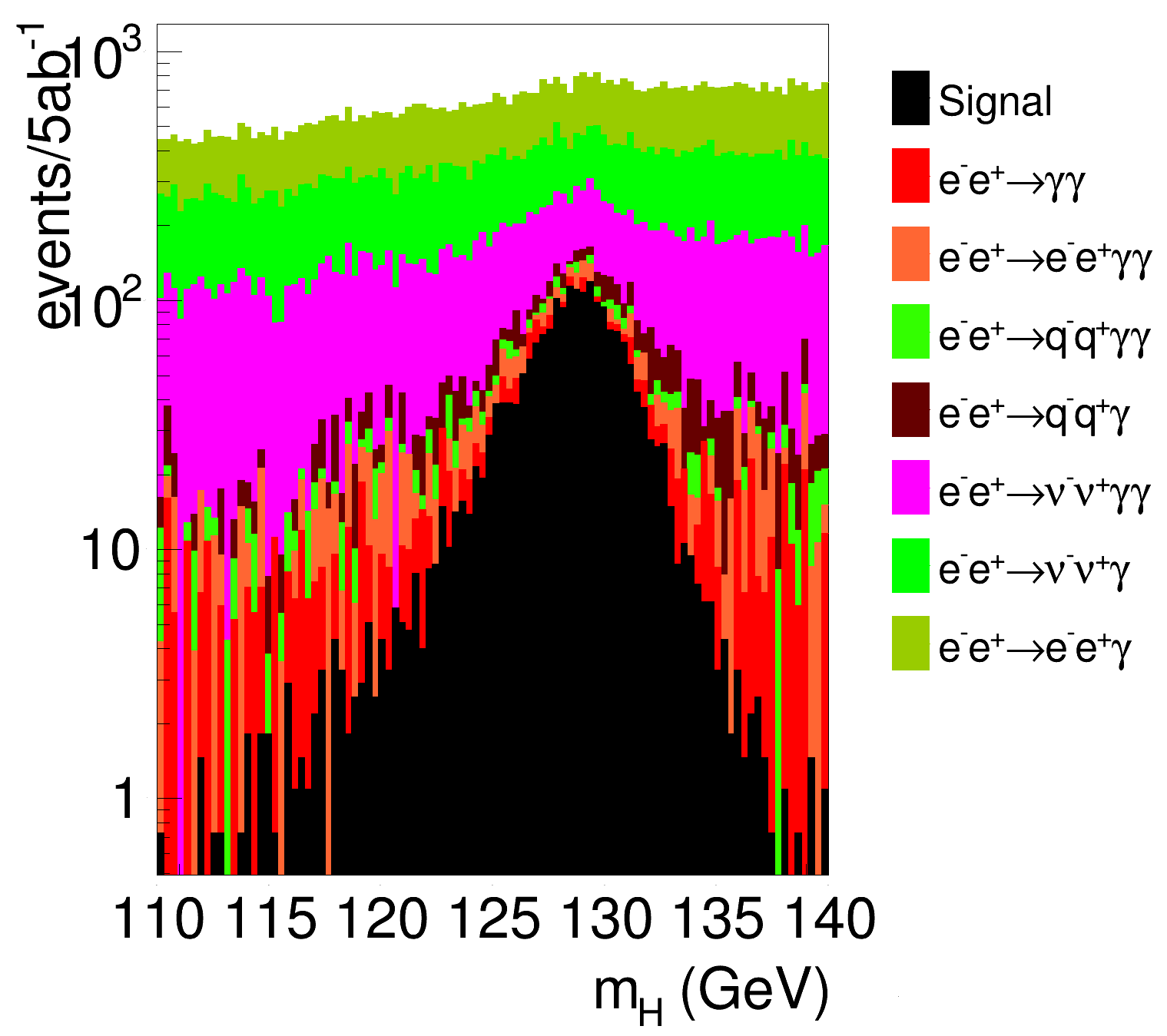}
        \begin{textblock}{3}(4.3,-4.35)
        \textbf{work in progress}
        \end{textblock}
\caption{\label{fig-stackMV}Stacked histograms of Higgs candidate mass distributions for signal and background after MVA selection.}
\end{figure}

The first step of the event selection is to find two isolated photons with transverse momenta, $p_{T}$, greater than 15 GeV. A photon is considered isolated if the energy of the reconstructed particles in a 14 mrad cone around the photon is less than 20 GeV. An event will be preselected if:

\begin{itemize}
\item The reconstructed di-photon invariant mass is in the range from 110 GeV to 140 GeV, corresponding to the Higgs mass window,
\item The reconstructed di-photon energy is in the range between 100 GeV and 1000 GeV,
\item The reconstructed di-photon transverse momentum is in the range between 20 GeV and 600 GeV.
\end{itemize}
The preselection is optimized to suppress high cross-section backgrounds like $ e^{+}e^{+}\rightarrow e^{+}e^{-}\gamma $ and $e^{+}e^{-}\rightarrow e^{+}e^{-}\gamma\gamma $ and $e^{+}e^{-}\rightarrow q\bar{q}\gamma$ . Preselection efficiency for the signal is 70\%, while background is reduced by a factor of 10.  After the preselection, multivariate based selection employing Boosted Decision Tree Gradient (BDTG) classifier is applied for further separation of signal and background events using their kinematic properties. The BDTG method is trained on twelve sensitive observables, among others energy and transverse momentum of di-photon system and of individual photons, polar angle distributions and depositions in electromagnetic and hadronic calorimeters. The Multi Variate Analysis (MVA) selection is optimized in a way to maximize the statistical significance. The MVA selection efficiency is 60\%, while the overall signal efficiency is 42\%. The di-photon invariant mass distribution after the MVA is illustrated for signal and background in  \cref{fig-stackMV}. 

\subsection{Pseudo-experiment}
\label{sec:basics}

In order to measure the $ BR (H\rightarrow\gamma\gamma)$, the number of selected signal events $N_{s}$ has to be known. It is determined by fitting the pseudo-data di-photon invariant mass distribution with the pre-determined probability density functions (PDFs) describing the signal and irreducible background. Such a fit is consider as a pseudo-experiment. PDF functions are determined from simulation, describing the shapes of di-photon invariant mass distribution for signal and background.  Figure 4 (a, b and c) gives PDF function of signal, background and an example of a pseudo-experiment, respectively.

\begin{figure}[h]
  \centering
  \begin{subfigure}[b]{0.3\textwidth}
    \includegraphics[width=\textwidth]{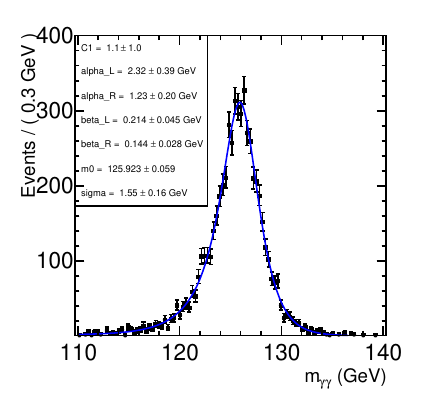}
        \begin{textblock}{}(2.3,-2.25)
        \textbf{work in progress}
        \end{textblock}
    \caption{}
    \label{fig:signalfit}
  \end{subfigure}
  \hfill
  \begin{subfigure}[b]{0.33\textwidth}
    \includegraphics[width=\textwidth]{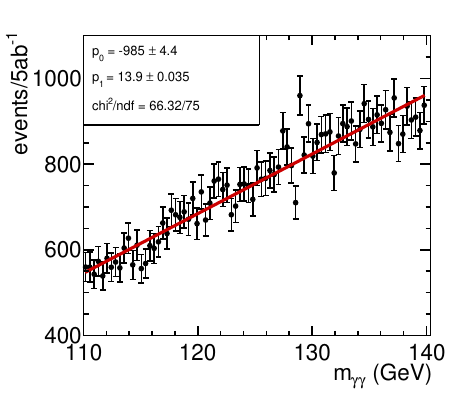} 
            \begin{textblock}{3}(1.5,-0.75)
        \textbf{work in progress}
        \end{textblock}
    \caption{}
  \label{fig:bckfit}
  \end{subfigure}
    \begin{subfigure}[b]{0.33\textwidth}
      \includegraphics[width=\textwidth]{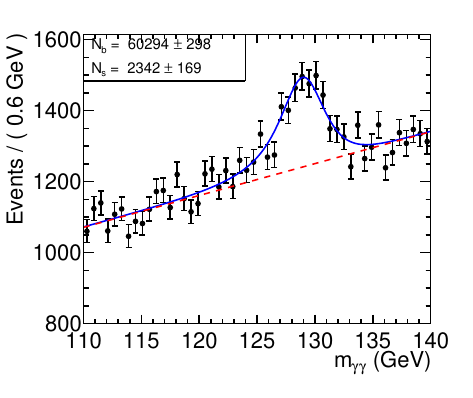}
              \begin{textblock}{3}(1.5,-0.75)
        \textbf{work in progress}
        \end{textblock}
      \caption{}
    \label{fig:combined}
    \end{subfigure}
\label{fig:example_analysis}
\caption{The di-photon mass distribution for (a) the signal, (b) the background and (c) an example of a pseudo-experiment. The lines are the result of a fit to PDFs (see text).}
\end{figure}

\subsection{Uncertainty of the measurement}
\label{sec:basics}
\begin{figure}[h]

\centering

\includegraphics[width=0.6\textwidth]{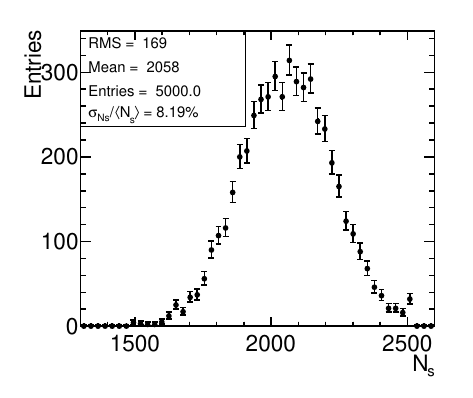}
        \begin{textblock}{3}(6.5,-3.9)
        \textbf{work in progress}
        \end{textblock}

\caption{\label{fig-toymc}Stacked histograms of Higgs candidate mass distributions for signal and background after MVA selection.}

\end{figure}

In order to estimate the statistical uncertainty of the measurement, 5000 pseudo-experiments are performed. The RMS of the resulting pull distribution is taken as the estimate of the statistical uncertainty of the measurement. From \cref{fig-toymc} the statistical uncertainty on the extracted number of signal events is therefore 8.2\%.  

\noindent Several sources of the systematic uncertainty of the measurement are considered like the uncertainty of the single photon identification efficiency and photon energy measurement, and the uncertainty of the integrated luminosity measurement and of the luminosity spectrum reconstruction. The dominant contribution of ~2\% comes from the 1\% uncertainty of the photon reconstruction efficiency. All other contributions are of order of several permille. The resulting systematic uncertainty of 2.4\% is smaller than the statistical one.

\section{Summary}
\label{sec:basics}
In a full simulation of the experimental measurement of the $BR (H\rightarrow\gamma\gamma)$ at 3 TeV CLIC, a statistical precision of 8.2\% is determined from 5000 pseudo-experiments, assuming $5 ab^{-1} $of integrated luminosity with unpolarized beams. By using 80\% beam polarisation and increasing the Higgs production cross-section, the statistical precision may improve by a factor of $\sqrt{1.48}$ . The total systematic uncertainty of 2.4\% is dominated by the uncertainty of photon identification efficiency.

\printbibliography[title=References]

\end{document}